\documentclass{iopart}
\usepackage{graphicx}
\usepackage{graphicx,epsf}

\begin{document}
\title{Spin-Polarized STM for a Kondo adatom}

\author{A. C. Seridonio,$^{1,2}$ F. M. Souza$^{1}$ and I. A. Shelykh$^{1,3}$}

\address{$^{1}$ICCMP - International Center for Condensed Matter Physics,
Universidade de Bras\'{\i}lia, 04513, Bras\'{\i}lia, DF, Brazil\\
 $^{2}$Instituto de F\'{\i}sica, Universidade Federal Fluminense, 24310-246, Niteroi, RJ, Brazil\\
 $^{3}$Science Department, University of Iceland, Dunhaga 3, IS-107, Reykjavik, Iceland}
\ead{fmsouza@unb.br}

\date{\today{}}

\begin{abstract}
We investigate the bias dependence of the tunneling conductance
between a spin-polarized (SP) scanning tunneling microscope (STM)
tip and the surface conduction states of a normal metal with a Kondo
adatom. Quantum interference  between tip-host metal and
tip-adatom-host metal conduction paths is studied in the full range of the Fano parameter $q$.
The spin-polarized STM gives rise to a splitting of the Kondo peak and
asymmetry in the zero-bias anomaly depending on the lateral
tip-adatom distance. For increasing lateral distances, the Kondo
peak-splitting shows a strong suppression and the spin-polarized
conductance exhibits the standard Fano-Kondo profile.
\end{abstract}


\maketitle

\section{Introduction}

The Kondo effect is an antiferromagnetic screening of a localized
magnetic moment by the host metallic electrons below a
characteristic Kondo temperature $T_K$, which results in the
appearance of an additional peak in the system's density of states
pinned to the Fermi energy. Being first observed during studies of
transport properties of bulk diluted magnetic alloys \cite{key-121},
the Kondo effect was later on shown to affect the conductance of
single quantum dots (QDs) \cite{key-122,key-123}, arrays of QDs
 \cite{array1,array2} and possibly quantum point
contacts \cite{QPC1,QPC2}. In the emerging field of
spintronics \cite{key-124}, the coupling of a single QD to
ferromagnetic leads can shed more light on fundamental aspects of
the Kondo physics and provide a basis for a design of novel
spintronic
composants \cite{pz02,key-127,jm03,key-128,jm05,yu05,rs06,ferro,anp04}.

The Scanning Tunneling Microscope (STM) is widely used for investigation of
the Kondo effect at surfaces of normal metals with adsorbed magnetic impurities (Kondo adatoms) \cite{key-115,hcm00,key-116,key-117,key-118,key-119,new1,new2,ou00}.
The possible examples are
individual Co atoms at Au(111), Cu(100) or Cu(111) surfaces \cite{key-115,hcm00,key-116,key-117}, and
cobalt carbonyl Co(CO)$_n$ complexes or manganese phthalocyanine
(MnPc) molecules on top of Pb islands \cite{key-118,key-119}.

\begin{figure}
 \includegraphics[%
   width=1.0\columnwidth,
   height=0.30\linewidth,
   keepaspectratio]{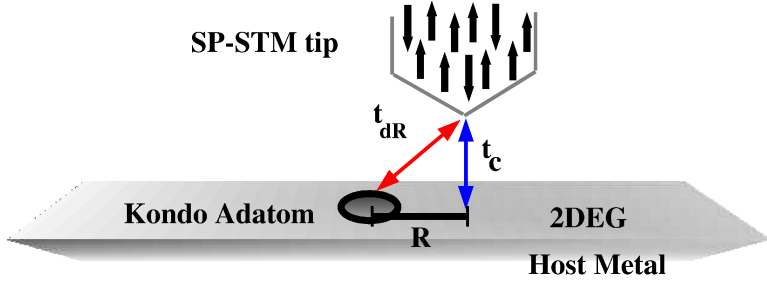}
   \caption{\label{fig1} Setup of the SP-STM and the host metal with the Kondo adatom. The interference between the channels $t_{dR}$ and $t_{c}$ exhibits a spin-polarized Fano-Kondo profile for the conductance.}
 \end{figure}

In the present work we focus on the Kondo regime for a system
containing a ferromagnetic STM tip and a single Kondo adatom on a
metallic surface {(}figure \ref{fig1}{)}. In this system the
conductance becomes spin dependent, due to the ferromagnetic (FM)
tip, which leads to a Zeeman splitting in the Kondo adatom density
of states (DOS). Particular attention is payed to the interplay
between Kondo effect, quantum interference of two possible tunneling
paths (tip-adatom-host and tip-host) and the ferromagnetism of the
tip.

Our study differs from previous work in the same field \cite{key-134} mainly in the three aspects described below.

\emph{First}. We present new results for the tunneling conductance
as a function of the bias for different lateral tip-adatom
separations R in the Kondo regime. This is an actual task indeed, as
in experiments with unpolarized systems, R was shown to strongly
affect the conductance \cite{key-115,hcm00}. For ferromagnetic tips,
however, correspondent results are lacking.

\emph{Second}. We discuss the small, large and intermediate cases
for the Fano parameter $q$ \cite{qdefinition,key-129,key-129b,key-129c,key-129d}. We show that the three cases above give
different conductance patterns. Experimentally,  the
large and small $q$ limits are relevant and have been realized for
unpolarized conduction bands. For example, in reference \cite{key-117}
it was shown that $q$ alternates from intermediate ($q \approx 1.13$)
to small ($q \approx 0.18$) values by changing the Cu crystal surface (Cu(100) and Cu(111)),
in which a Co adatom is deposited. The conductance profile for small $q$ limit
can also be experimentally observed in quantum corrals \cite{hcm00}. Alternatively, manipulating the molecular
structure of the magnetic impurity on the Cu(100) surface, it is
possible to switch between the intermediate and large $q$
limits \cite{key-118,key-119}.

\emph{Third}. Instead of using a Lorentzian approximation for
the Kondo peak, we apply the well established Doniach-Sunjic
formula \cite{key-130,key-132}, whose validity is supported by
numerical calculations based on both numerical renormalization
group \cite{key-130,key-132}, quantum Monte Carlo
simulations \cite{rns90} and by the recent success in fitting
experimental data \cite{key-119}.

Our results show that for a tip situated right above the Kondo
adatom (\textit{R=0}) an asymmetric zero-bias anomalies appear,
which are revealed as resonances and anti-resonances in the
conductance in the limits of large and small $q$, respectively. For
$q\approx 1$, the conductance demonstrates a pronounced plateau in
the region of small biases ($eV\approx 0$). The increase of $R$
leads to a suppression of the tip-induced adatom's Zeeman splitting,
thus resulting in a conductance pattern that resembles experimental
data for unpolarized systems \cite{key-115}. Finally, we verify that
the Kondo peak splitting strongly depends on the asymmetry between
the tip-adatom and adatom-host tunnelings.

The paper is organized as follow. In section 1 we present the model
adopted to describe the system under study. In section 2 we derive an
expression for the spin-resolved conductance based on a perturbative
expansion for the tunneling Hamiltonian. In section 3 we discuss the
numerical results. Conclusions are present in section 4.

\section{The Model}

The description of the metallic host and its interaction
with the adatom is performed within the framework of the single impurity Anderson model \cite{key-120}
with a half-filled noninteracting conduction band,

\begin{eqnarray}
H_{A} & = & \sum_{\sigma}\int\varepsilon c_{\varepsilon\sigma}^{\dagger}c_{\varepsilon\sigma}d\varepsilon+\sum_{\sigma}\varepsilon_{d\sigma}d_{\sigma}^{\dagger}d_{\sigma}+Un_{d\uparrow}n_{d\downarrow}\nonumber \\
 & + & \sqrt{\frac{\Gamma}{\pi}}\sum_{\sigma}\int d\varepsilon\left(c_{\varepsilon\sigma}^{\dagger}d_{\sigma}+H.c.\right),\label{eq:1}\end{eqnarray}
where all energies are measured from the Fermi level coinciding with the center of the band $\left(\varepsilon_F=0\right)$ and
extend from $-D$ to $D$. The operator
\begin{equation}
c_{\varepsilon\sigma}=\rho_{0}^{-1/2}\sum_{\vec{k}}c_{\vec{k}\sigma}\delta(\varepsilon-\varepsilon_{k}),\label{eq:2}\end{equation}
corresponds to a surface conduction state of the host metal with an
energy-independent density of states per spin $\rho_{0}$ .

The first term in  equation (\ref{eq:1}) describes a two dimensional
electron gas (2DEG) on the surface of a host metal. The second and the
third correspond to the adatom, which is characterized by a single
particle orbital energy $\varepsilon_{d\sigma}$ and Coulomb
repulsion $U$. The last term, proportional to $\sqrt{\Gamma},$
describes the coupling between the adatom and the host metal
conduction states, thus introducing the broadenings of the adatom's
resonances at the energies $\varepsilon_{d\sigma}$ and
$\varepsilon_{d\sigma}+U.$ In the Kondo regime ($T \ll T_K,
\varepsilon_{d\sigma}<\varepsilon_F,
\varepsilon_{d\sigma}+U>\varepsilon_F$, $\Gamma \ll
|\varepsilon_{d\sigma}|$, $\varepsilon_{d \sigma}+U$) an additional
peak in the density of states having a half-width
$\Gamma_{K}=k_{B}T_{K}$, ($k_B$ is the Boltzmann constant and $T_K$
the Kondo temperature) \cite{TKdefinition}  appears exactly at the
Fermi level.

The total STM Hamiltonian reads

\begin{equation}
H_{STM}=H_{A}+H_{tip}+H_{tun},\label{eq:3}\end{equation} where
$H_{tip}$ corresponds to free electrons in the tip,

\begin{equation}
H_{tip}=\sum_{\vec{p}\sigma}(\varepsilon_{\vec{p}}+eV)a_{\vec{p}\sigma}^{\dagger}a_{\vec{p}\sigma},\label{eq:4}\end{equation}
with the operators $a_{\vec{p}\sigma}$ describing the bulk conduction
states, the bias is $eV$ and

\begin{equation}
H_{tun}=\sum_{\vec{p}\sigma}\left[t_{c}^{\sigma}a_{\vec{p}\sigma}^{\dagger}B_{\sigma}(\vec{R})+H.c.\right]\label{eq:5}\end{equation}
is the tunneling Hamiltonian that connects the tip with the host metal via the operator

\begin{equation}
B_{\sigma}(\vec{R})=\left[\int{\tilde{N}_{\varepsilon}^{-1}}\tilde{C}_{\varepsilon\sigma}d\varepsilon+q_{R}^{\sigma}\sqrt{\pi\Gamma\rho_{0}}d_{\sigma}\right].\label{eq:6}\end{equation}
The normalization factor in equation (\ref{eq:6}) reads
\begin{equation}
\tilde{N}_{\varepsilon}=\left[{\sum_{\vec{k}}|\varphi_{\vec{k}}(\vec{R})|^{2}\delta(\varepsilon-\varepsilon_{k})}\right]^{-1/2},\label{eq:8}
\end{equation}
with $\varphi_{\vec{k}}(\vec{R})\sim e^{i\vec{k}\vec{R}}$ being a wave function of the host conduction electron.

The first term in equation (\ref{eq:6}) containing the operator
\begin{equation}
\tilde{C}_{\varepsilon\sigma}=\tilde{N}_{\varepsilon}{\sum_{\vec{k}}\varphi_{\vec{k}}(\vec{R})c_{\vec{k}\sigma}\delta(\varepsilon-\epsilon_{k})},\label{eq:7}\end{equation}
hybridizes the conduction states of the tip with the surface of the
host at a displaced lateral position $\vec{R}$ from the adatom.

The second term describes the tunneling between the tip and localized adatom's level characterized by a spin-dependent  parameter (Fano factor) \cite{key-41,key-42}
\begin{equation}
q_{R}^{\sigma}=\left(\pi\Gamma\rho_{0}\right)^{-1/2}\left(t_{dR}^{\sigma}/t_{c}^{\sigma}\right),\label{eq:8.b}
\end{equation}
defined as a ratio between the couplings $t_{dR}^{\sigma}$ of the
tip-adatom and $t_{c}^{\sigma}$ of the tip-host metal.

The Fano factor monitors the competition between the tunneling
channels in the system. It vanishes with the increase of the
tip-adatom lateral distance $R$, which can be modeled by an
exponentially decaying function \cite{key-42}

\begin{equation}
q_{R}^{\sigma}=q_{R=0}^{\sigma}e^{-k_{F}R}.\label{eq:8.c}
\end{equation}

The decay of the Fano parameter with the increasing of the lateral
tip-adatom distance was already experimentally explored in a system
with a Co adatom on a Cu surface \cite{key-117}.

In the present work we consider a ferromagnetic tip with a spin-dependent density of states given by
\begin{equation}
\rho_{tip}^{\sigma}=\rho_{0}\left[1+\sigma P_{tip}\right],\label{eq:8d}
\end{equation}
where $\sigma=+$ or $-$ for spins $\uparrow$ or $\downarrow$,
respectively, $P_{tip}$ is a polarization degree of the tip. The
inequality between spin-up and spin-down populations in the tip
$\rho_{tip}^{\uparrow}> \rho_{tip}^{\downarrow}$ introduces an
asymmetry in the splitting of the zero-bias anomaly as we will see
in Sec. IV. The density of states for the unpolarized tip
$\rho_{0}=\rho_{tip}(\varepsilon_F)$ is assumed to be equal to the
density of states of the host metal for simplicity.

\section{Methodology}

We calculate the tunneling conductance of the system treating the
coupling between tip and host metal $(H_{tun})$ as a perturbation.
Within a second order perturbation scheme, the formula for the
conductance \cite{key-41} reads
\begin{equation}
G=\left(e^{2}/h\right)\sum_{\sigma}\int T_{\sigma}(\varepsilon,T,R,q_{R}^{\sigma})\left[-\frac{\partial}{\partial\varepsilon}f(\varepsilon-eV)\right]d\varepsilon,\label{eq:9}
\end{equation}
 where
\begin{equation}
T_{\sigma}(\varepsilon,T,R,q_{R}^{\sigma})=T_{o\sigma}\left\{ 1+\left|q_{R}^{\sigma}\right|^{2}\right\} \left(\rho_{tip}^{\sigma}/\rho_{0}\right)\left(\rho_{LDOS}^{\sigma}/\rho_{0}\right),\label{eq:10}
\end{equation}
is an effective transmission coefficient with
\begin{equation}
T_{o\sigma}=\left(2\pi\rho_{0}t_{c}^{\sigma}\right)^{2}.\label{eq:10b}
\end{equation}
The local density of states (LDOS) appearing in equation (\ref{eq:10}) is defined as
\begin{equation}
\rho_{LDOS}^{\sigma}=-\frac{1}{\pi} \frac{\Im \langle \langle B_{\sigma}(\vec{R})|B_{\sigma}^{\dagger}(\vec{R})\rangle \rangle_{\varepsilon}}{1+\left|q_{R}^{\sigma}\right|^{2}},\label{eq:11}
\end{equation}
with $\langle \langle B_{\sigma}(\vec{R})|B_{\sigma}^{\dagger}(\vec{R}) \rangle \rangle_{\varepsilon}$
being the retarded Green function thermally averaged over the
eigenstates of the Hamiltonian (\ref{eq:1}).

Looking at equation (\ref{eq:11}), one sees that the LDOS depends on  non-orthonormal fermionic operators
\begin{equation}
\{\tilde{C}_{\varepsilon\sigma}^{\dagger},c_{\varepsilon^{\prime}\sigma}\}={\frac{\tilde{N}_{\varepsilon}}{\sqrt{\rho_{0}}}}\delta(\varepsilon-\varepsilon^{\prime})F_{\vec{R}}(\varepsilon),\label{eq:orto_01}
\end{equation}
where the spatial function $F_{\vec{R}}(\varepsilon)$ is given by
\begin{equation}
F_{\vec{R}}(\varepsilon)=\sum_{\vec{k}}\varphi_{\vec{k}}(\vec{R})\delta(\varepsilon-\varepsilon_{k})\label{eq:orto_02}=\rho_0J_0(k(\varepsilon)R),
\end{equation}
with $J_0$ being the 0-th order Bessel function.

The operator (\ref{eq:6}) can be expressed in terms of fermionic
operators orthonormal to $c_{\varepsilon\sigma}$ by introducing
\begin{equation}
\tilde{c}_{\varepsilon\sigma}=N_{0}\left(\tilde{C}_{\varepsilon\sigma}-{\frac{\tilde{N}_{0}}{\sqrt{\rho_{0}}}}F_{\vec{R}}(0)c_{\varepsilon\sigma}\right),\label{eq:orto_03}\end{equation}
with a normalization factor evaluated at the Fermi level,
\begin{equation}
N_{0}=\left\{ 1-\left|{\frac{\tilde{N}_{0}}{\sqrt{\rho_{0}}}}F_{\vec{R}}(0)\right|^{2}\right\} ^{-1/2}.\label{eq:orto_04}\end{equation}
This leads to the following expression for $B_{\sigma}(\vec{R})$,
\begin{eqnarray}
B_{\sigma}(\vec{R})\nonumber \\
= & \left[{\frac{\int d\varepsilon\tilde{c}_{\varepsilon\sigma}}{\tilde{N}_{0}N_{0}}}+\sqrt{2}\frac{F_{\vec{R}}(0)}{\rho_{0}}f_{0\sigma}+q_{R}^{\sigma}\sqrt{\pi\Gamma\rho_{0}}d_{\sigma}\right],\nonumber \\
\label{eq:orto_05}
\end{eqnarray}
where we introduce an operator
\begin{equation}
f_{0\sigma}=\sqrt{\frac{\rho_{0}}{2}}\int c_{\varepsilon\sigma}d\varepsilon,\label{eq:orto_06}
\end{equation}
which describes a conduction state centered at the adatom site.

As the Kondo effect occurs at low temperatures $T \ll T_{K}$ and the
tip bias is usually much smaller then the bandwidth, $eV \ll D$, we evaluate equation (\ref{eq:9}) at $T=0$, thus resulting in
\begin{equation}
G(eV,T \ll T_{K},R)=\sum_{\sigma}G_{max}^{\sigma}\left(\rho_{tip}^{\sigma}/\rho_{0}\right)\left(\rho_{LDOS}^{\sigma}/\rho_{0}\right)\label{eq:18}
\end{equation}
 where
\begin{equation}
G_{max}^{\sigma}=\left(e^{2}/h\right)T_{o\sigma}\left\{ 1+\left|q_{R}^{\sigma}\right|^{2}\right\} \label{eq:18b}\end{equation}
and
\begin{eqnarray}
\rho_{LDOS}^{\sigma}/\rho_{0} & = & \left[1-J_{0}^{2}\left(k_{F}R\right)\right]\cos^{2}\delta_{q_{R}}^{\sigma}+\sin^{2}\delta_{q_{R}}^{\sigma}\sin^{2}\delta_{eV}^{\sigma}\nonumber \\
 & + & 2J_{0}\left(k_{F}R\right)\sin\delta_{q_{R}}^{\sigma}\cos\delta_{q_{R}}^{\sigma}\sin\delta_{eV}^{\sigma}\cos\delta_{eV}^{\sigma}\nonumber \\
 & + & J_{0}^{2}\left(k_{F}R\right)\cos^{2}\delta_{q_{R}}^{\sigma}\cos^{2}\delta_{eV}^{\sigma}.\label{eq:19}
\end{eqnarray}
In the calculation of the above expression we used the Green
function identities for zero temperature Anderson model \cite{key-120}. The spin dependent Fano factor phase shift
$\delta_{q_{R}}^{\sigma}$ is defined as
\begin{equation}
\tan\delta_{q_{R}}^{\sigma}\equiv\left|q_{R}^{\sigma}\right|.\label{eq:19b}\end{equation}

In equation (\ref{eq:19}) the terms proportional
to $\sin^{2}\delta_{eV}^{\sigma}$ and $\cos^{2}\delta_{eV}^{\sigma}$
come from the direct tunneling paths tip-adatom-host and tip-host, respectively. The interference between then is given
by the mixture term proportional to $\sin\delta_{eV}^{\sigma}\cos\delta_{eV}^{\sigma}.$

The spin dependent phase shift $\delta_{\varepsilon}^{\sigma}$  for
the conduction states can be determined from the Doniach-Sunjic spectral
density \cite{key-130,key-132}
\begin{equation}
\rho_{dd}^{\sigma}(eV)=\frac{1}{\pi\Gamma}\Re\left[\frac{i\Gamma_{K}}{\left(eV+\sigma\tilde{\Delta}\right)+i\Gamma_{K}}\right]^{\frac{1}{2}}=\frac{1}{\pi\Gamma}\sin^{2}\delta_{eV}^{\sigma},\label{eq:20}
\end{equation}
where $\tilde{\Delta}$ gives the Kondo peak splitting, which is related to the adatom energy level as
\begin{equation}
 \varepsilon_{d \sigma}= \varepsilon_d^0 - \sigma \frac{\tilde{\Delta}}{2},
\end{equation}
where $\varepsilon_d^0$ is the adatom level without magnetic field.
According to ``poor man's'' scaling \cite{jm03}, such splitting can
be estimated as \cite{Deltatio}
\begin{eqnarray}
\tilde{\Delta}&=&\frac{\Gamma_{tip}^{\uparrow}+\Gamma_{tip}^{\downarrow}}{2\pi}P_{tip}\ln(D/U)\nonumber \\ &=&\left[\frac{\Gamma_{tip}^{0}}{\pi}\ln(D/U)\right]P_{tip}\exp\left(-2k_FR\right),\label{eq:20.b.0}
\end{eqnarray}
where
$\Gamma_{tip}^{0}=\pi\left|t_{dR=0}^{\sigma}\right|^{2}\rho_{0}$
gives the local coupling between the Kondo adatom and an unpolarized
tip. We use the expression
\begin{equation}
 \Gamma_{tip}^{\sigma}=\Gamma_{tip}^{0}\left[1+\sigma P_{tip}\right]\exp\left(-2k_{F}R\right),
\end{equation}
to account for both spin and spatial dependencies of the tip-adatom coupling.

\section{Results}

For numerical analysis we adopt the following set of model parameters:
$\varepsilon_d^0=-0.9$eV, $\Gamma=0.2$eV, $U=2.9$eV, $D=5.5$eV,
$T_{K}=50$K and $k_F=0.189${\AA}$^{-1}$ \cite{ou00,key-134}. We
consider the cases of large, small and intermediate Fano ratio
values. For each of these cases we analyze the dependence of the
conductance on tip-adatom lateral distance $R$. Additionally,
 for large $q$ limit we consider how the conductance depends on
the asymmetry between tip-adatom and adatom-host couplings.

\begin{figure}
 \includegraphics[%
   width=2.5\columnwidth,
   height=0.70\linewidth,
   keepaspectratio]{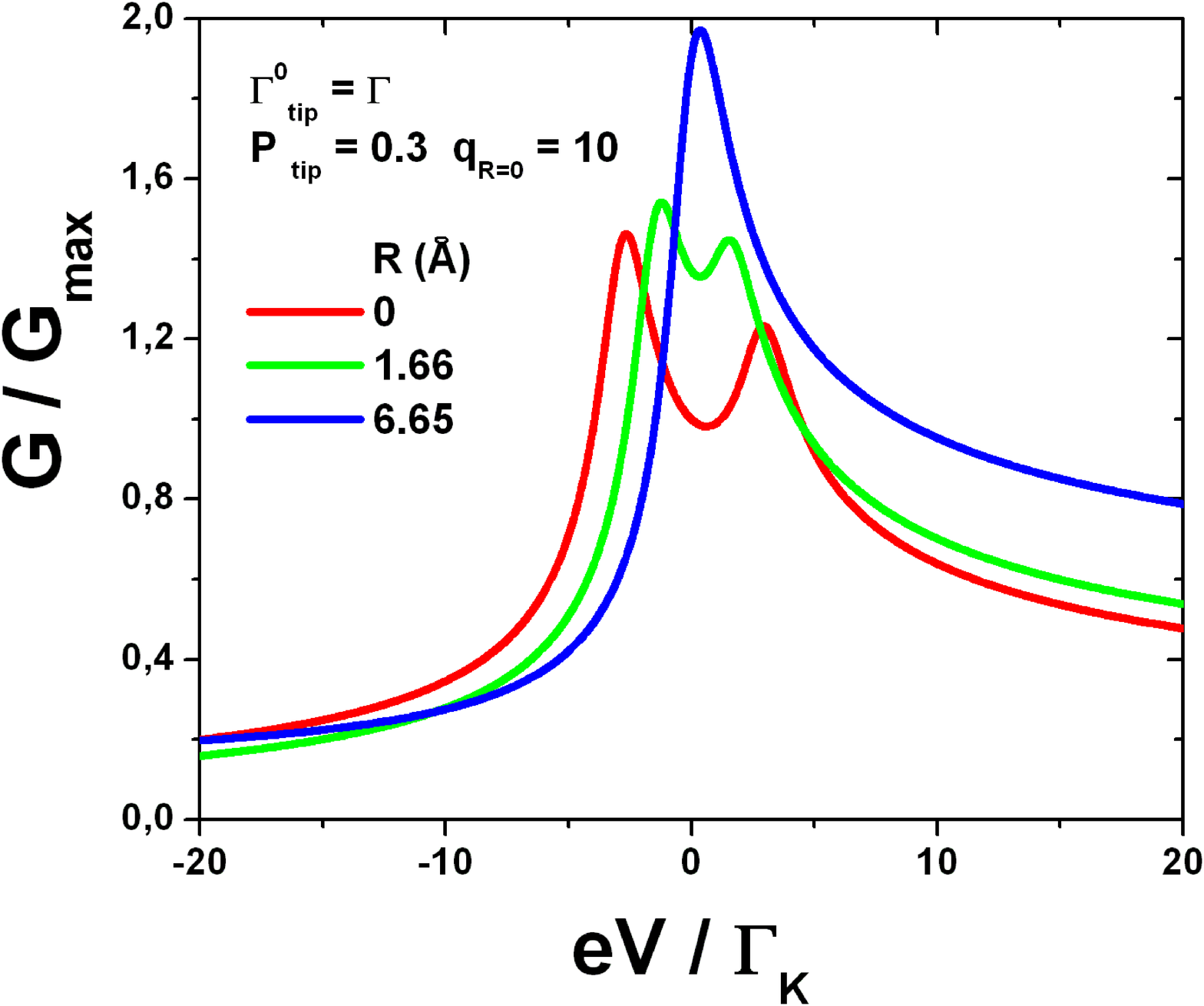}
  \caption{\label{fig2} Conductance $G/G_{max}$ as a function of the tip bias scaled in units of Kondo resonance half-width $eV/\Gamma_{K}$, for $q_{R=0}=10$ at three different tip lateral positions with symmetric potential barriers.}
 \end{figure}

\begin{figure}[h]
 \includegraphics[%
   width=2.5\columnwidth,
   height=0.70\linewidth,
   keepaspectratio]{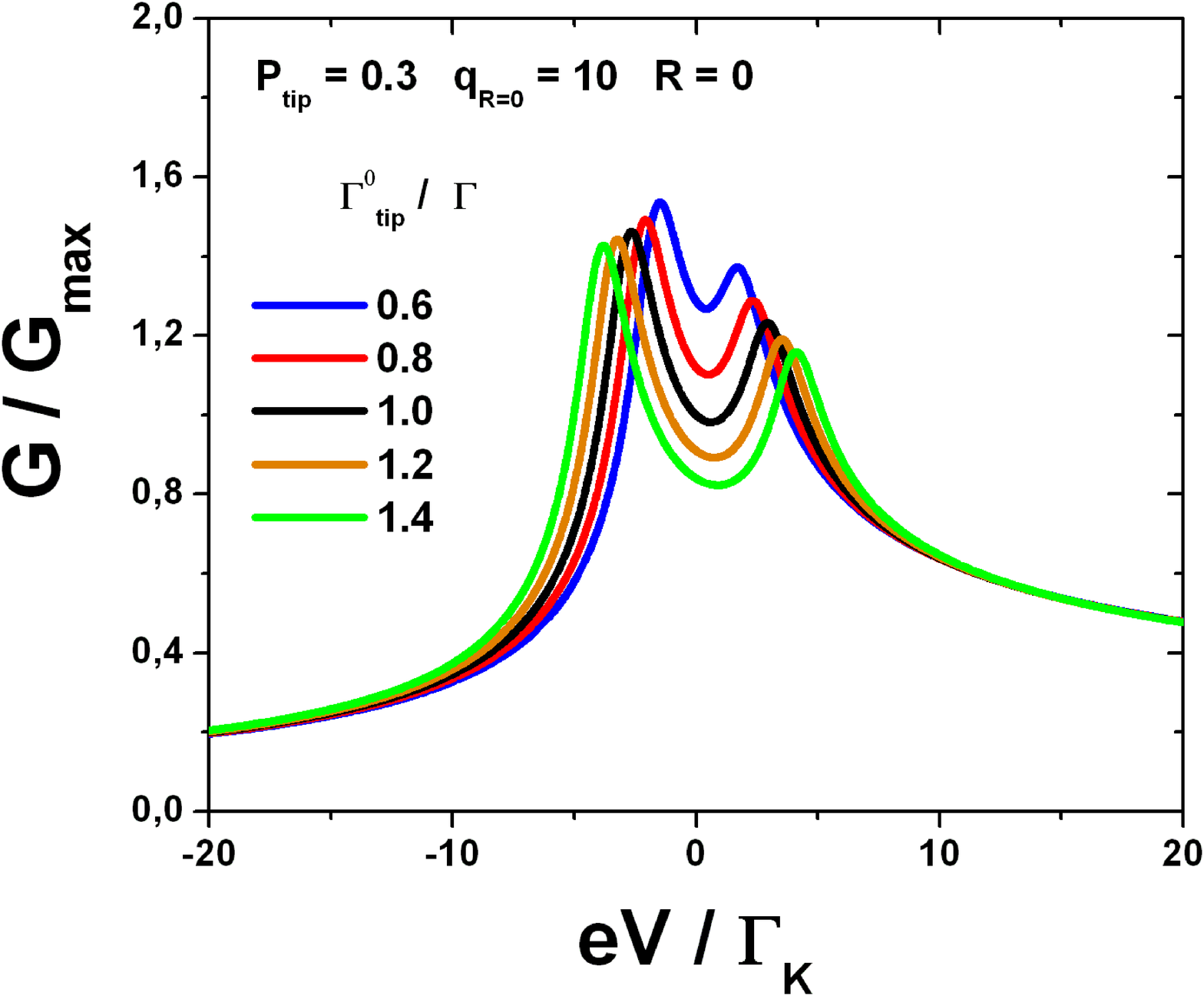}
  \caption{\label{fig3} Conductance $G/G_{max}$ at the origin as a function of the tip bias scaled in units of Kondo resonance half-width $eV/\Gamma_{K}$, for $q_{R=0}=10$  and different potential barriers.}
 \end{figure}

\subsection{Large tip-adatom coupling ($q_{R=0}=10$)}

The large $q$ limit has been achieved in experiments with STM tip by
employing magnetic molecules as Kondo adatoms \cite{key-118,key-119}.
In this limit the host metal conduction electrons tunnel towards the
tip preferably via the localized magnetic adatom state. For the tip
situated right above the adatom $(R=0)$, the conductance reveals an
asymmetric splitting of the zero bias anomaly (figure (\ref{fig2})),
characterized by a pair of peaks at $eV=-\tilde{\Delta}$ and
$eV=\tilde{\Delta}$. Such asymmetry occurs due to the spin
polarization of the tip (equation (\ref{eq:8d})), the higher peak
corresponding to the majority spin up states, while the lower one to
the minority spin down states.

A similar asymmetric splitting of the Kondo peak was recently
observed in a quantum dot system coupled to two ferromagnetic Ni
electrodes \cite{ferro,anp04}. The large $q$ limit in the system we
consider resembles the standard case of a single dot in
between leads without lead-to-lead direct coupling (embedded geometry). A wealth of
theoretical works predict the spin splitting of the Kondo peak in a
quantum dot system coupled to two ferromagnetic leads \cite{key-127,jm03,rs06}.
In those works this splitting was tuned via the relative angle between the left
and the right lead magnetization \cite{rs06}, the leads polarization \cite{key-127}
and an external magnetic field \cite{jm03}. Here we show one alternative/additional
way to tune the spin splitting, by changing the tip-adatom separation (laterally or vertically) \cite{nonequilibriumKondo}.

Increasing the tip-adatom lateral distance, the Fano ratio decays according to equation (\ref{eq:8.c}) and the Zeeman splitting of the Kondo peak quenches (see equation (\ref{eq:20.b.0})). These effects can be seen at figure (\ref{fig2}), where the conductance is plotted for three different values of the tip-adatom lateral distance. About $R = 6.65${\AA}, the two resonances merge into a single peak, thus resulting in the standard Kondo resonance profile.

Not only the lateral tip-adatom separation can change the spin
splitting, but also the vertical tip-adatom distance. To mimic this
effect we change the ratio $\Gamma^0_{tip}/\Gamma$, thus introducing
an asymmetry between the tip-adatom ($\Gamma^0_{tip}$) and
adatom-host ($\Gamma$) tunneling rates. When the tip becomes
vertically closer to the adatom, we expect the increase of the
coupling parameter $\Gamma^0_{tip}$. Consequently, the splitting
between the peaks becomes more apparent (see figure (\ref{fig3})) which
can be understood as consequence of the enhancement of the local tip
magnetic field on the adatom due to the tip proximity.

\begin{figure}
 \includegraphics[%
   width=2.5\columnwidth,
   height=0.70\linewidth,
   keepaspectratio]{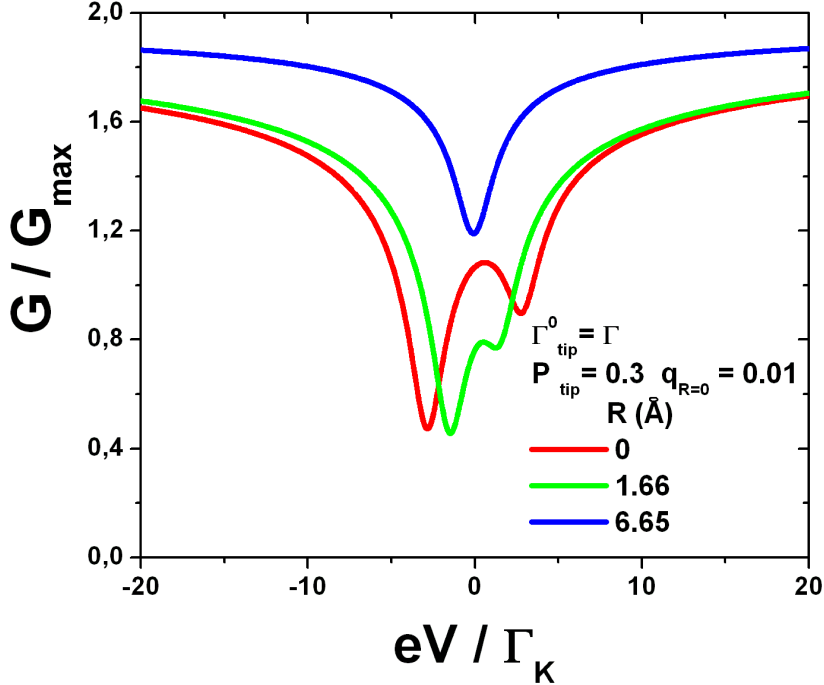}
  \caption{\label{fig4} Conductance $G/G_{max}$ as a function of the tip bias scaled in units of Kondo resonance half-width $eV/\Gamma_{K}$, for $q_{R=0}=0.01$ at three different tip lateral positions with symmetric potential barriers.}
 \end{figure}

\subsection{Small tip-adatom coupling ($q_{R=0}=0.01$)}

In the small coupling limit the conductance curves (figure (\ref{fig4})) display dips instead of peaks observed in the large
coupling limit (figures (\ref{fig2})-(\ref{fig3})). The appearance of
the dips at $eV=\pm\tilde{\Delta}$ is a consequence of a destructive
quantum interference between the channels tip-host and
tip-adatom-host. This can be easily seen from equations
(\ref{eq:19})-(\ref{eq:19b}) for $q \to 0$ and $R=0$ that the LDOS behaves as
\begin{equation}
 \rho_{LDOS}^{\sigma}/\rho_{0} = \cos^{2}\delta_{eV}^{\sigma},
\end{equation}
which is opposite to the case $q_{R=0} \to \infty$, where
\begin{equation}
\rho_{LDOS}^{\sigma}/\rho_{0} = \sin^{2}\delta_{eV}^{\sigma}.
\end{equation}

As in the case of large coupling, the increase of the tip-adatom
distance leads to a quenching of the anti-resonance splitting as it
can be seen comparing the curves corresponding to different values of $R$ at
figure (\ref{fig4}). At $R =6.65$\AA, the asymmetric zero-bias anomaly
for the dips disappears and only the standard single anti-resonance
profile is recovered. A single dip structure in the conductance for
small $q$ is verified for a system composed of Co on Cu(111)
surface \cite{hcm00,key-117}. It is valid to note that while the
large $q$ limit resembles the embedded geometry, the small $q$ limit
gives similar results to the T-shaped quantum dot (side-coupled
geometry) \cite{key-123,key-46,key-46b}.

\subsection{Intermediate tip-adatom coupling ($q_{R=0} = 1$)}

\begin{figure}
 \includegraphics[%
   width=2.5\columnwidth,
   height=0.70\linewidth,
   keepaspectratio]{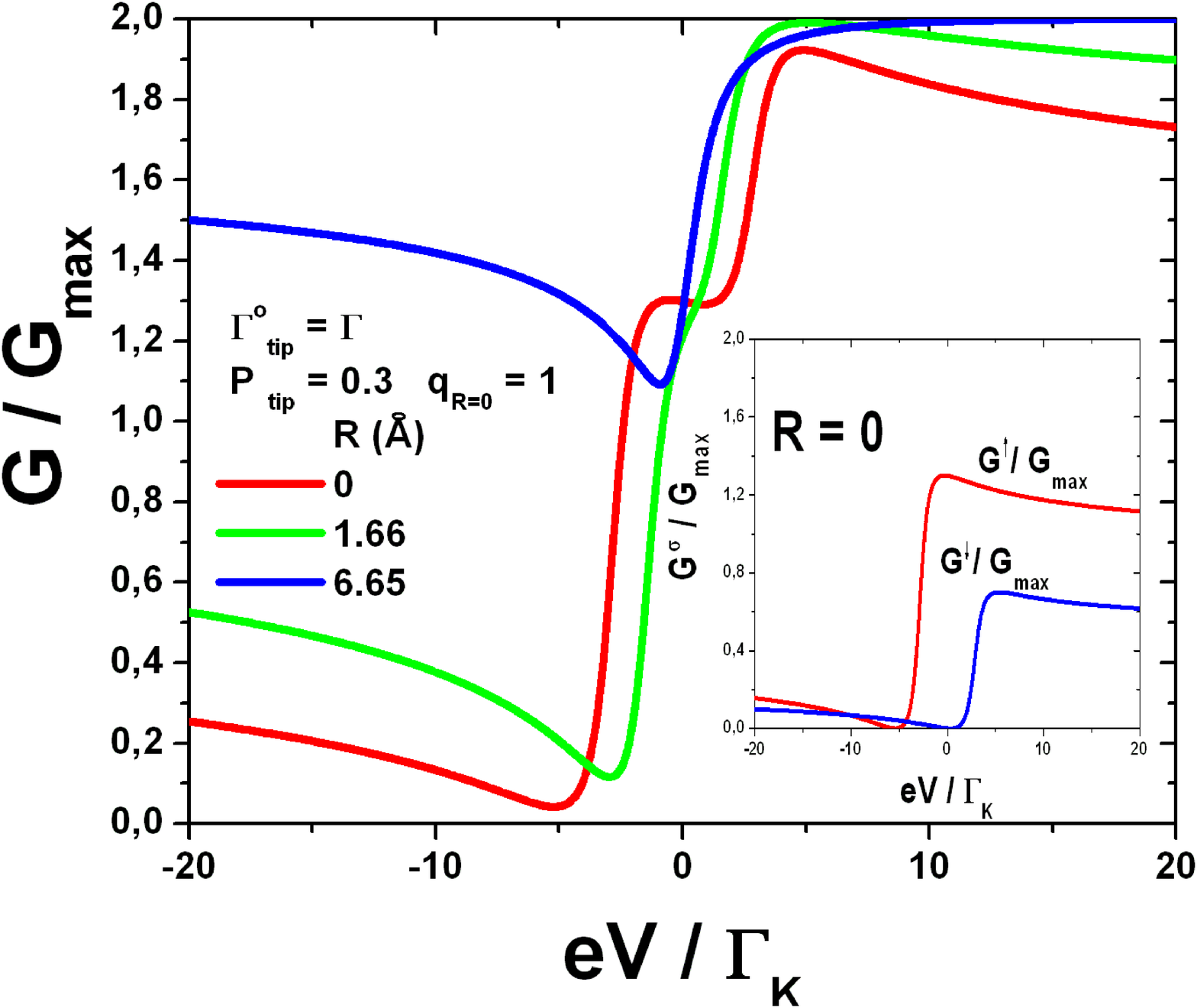}
  \caption{\label{fig5} Conductance $G/G_{max}$ as a function of the tip bias scaled in units of Kondo resonance half-width $eV/\Gamma_{K}$, for $q_{R=0}=1$ and three different tip lateral positions with symmetric potential barriers.}
 \end{figure}

The intermediate case for a normal tip is characterized by the well
known Fano-Kondo line shape \cite{key-115,key-117}. However, the
introduction of a ferromagnetic tip results in distinct Fano-Kondo
profiles for each spin component. The spin up profile is shifted
toward negative bias with an enhanced amplitude, while the spin down
case moves in the opposite direction being reduced in amplitude.
This is illustrated in the inset of figure (\ref{fig5}), where we show
$G^\uparrow/G_{max}$ and $G^\downarrow/G_{max}$.

The superposition of the spin-dependent Fano-Kondo profiles gives
rise to the appearance of a plateau around the Fermi level ($eV=0$)
in the total conductance ($G^\uparrow+G^\downarrow$) for $R=0$ (see
figure (\ref{fig5})). For large $R$ ($R = 6.65${\AA}), this plateau vanishes due to the
suppression of the adatom's Zeeman splitting and standard Fano-Kondo
results for the conductance with nonmagnetic tips are
recovered \cite{key-115,key-117}.

\section{Conclusions}

We derived a spin resolved tunneling conductance for a system of a
spin polarized STM tip with a single Kondo adatom on the surface of
a normal metallic host. The conductance dependence on the tip
bias was investigated for different tip-adatom lateral distances in
a wide range of the Fano parameter $q$, relevant for various
experimental configurations. We demonstrated that the Fano parameter
drastically affects the conductance pattern of the system. For large
values of $q$, we observe an asymmetric splitting of the Kondo
resonance which is suppressed with an increase of the tip-adatom
lateral distance. For small q, the behavior of the conductance is
opposite to those observed in the large $q$ regime- instead of a
splitted Kondo peak, one observes an asymmetrically splitted Kondo
dip. For the intermediate case we have shown that due to the
splitting of the spin resolved conductances ($G^\uparrow \neq
G^\downarrow$), the total conductance exhibits a plateau in the
region of small biases, which disappears for large enough values of
a tip-adatom lateral distance.

\section{Acknowledgments}

We thank L. N. Oliveira and M. Yoshida for stimulating discussions.
This work was supported by the Brazilian agencies IBEM and CAPES.


\section*{References}

\begin{thebibliography}{99}

\bibitem{key-121} Hewson A C, 1993 {\it The Kondo Problem to Heavy Fermions} (Cambridge University
Press, Cambridge)

\bibitem{key-122} Goldhaber-Gordon D, Shtrikman H, Mahalu D, Abusch-Magder D, Meirav U and Kastner M A 1998 {\it Nature} {\bf 391} 156

\bibitem{key-123} Sato M, Aikawa H, Sano A, Katsumoto S and Yie Y 2005 {\it Phys. Rev. Lett.}
{\bf 95} 066801

\bibitem{array1} Busser C A, Moreo A and Dagotto E 2004 {\it Phys. Rev.} B {\bf 70} 035402

\bibitem{array2} Lobos A M and Aligia A A 2006 {\it Phys. Rev.} B {\bf 74} 165417

\bibitem{QPC1} Sfigakis F, Ford C J, Pepper M, Kataoka M, Ritchie D A and Simmons M Y 2008
{\it Phys. Rev. Lett.} {\bf 100} 026807

\bibitem{QPC2} Tripathi V and Cooper N R 2008 {\it J. Phys: Condens. Matter} {\bf 20} 164215

\bibitem{key-124} For a review on spintronics see Zutic I, Fabian J, and Das Sarma S 2004 {\it Rev. Mod. Phys.} {\bf 76} 323



\bibitem{pz02} Zhang P, Xue Q K, Wang Y and Xie X C 2002 {\it Phys.
Rev. Lett.} {\bf 89} 286803

\bibitem{key-127} Martinek J, Sindel M, Borda L, Barna{\'s} J, K{\"o}nig J, Sch{\"o}n G and
von Delft J 2003 {\it Phys. Rev. Lett.} {\bf 91} 247202

\bibitem{jm03} Martinek J, Utsumi Y, Imamura H, Barna{\'s} J, Maekawa S, K{\"o}nig J and Sch{\"o}n G 2003
{\it Phys. Rev. Lett.} {\bf91} 127203

\bibitem{key-128} Mahn-Soo Choi, S{\'a}nchez D and L{\'o}pez R 2004 {\it Phys. Rev. Lett.} {\bf 92}
056601

\bibitem{jm05} Martinek J, Sindel M, Borda L, Barna{\'s} J, Bulla R, K{\"o}nig J, Sch{\"o}n G,
Maekawa S and von Delft J 2005 {\it Phys. Rev.} B {\bf 72} 121302(R)

\bibitem{yu05} Utsumi Y, Martinek J, Sch{\"o}n G, Imamura H and Maekawa S 2005 {\it Phys. Rev.} B {\bf 71}
245116

\bibitem{rs06} {\'S}wirkowicz R, Wilczy{\'n}ski M, Wawrzyniak M and Barna{\'s} J 2006
{\it Phys. Rev.} B {\bf 73} 193312

\bibitem{ferro} Hamaya K, Kitabatake M, Shibata K, Jung M, Kawamura M, Hirakawa K, Machida T and
Taniyama T 2007 {\it Appl. Phys. Lett.} {\bf 91} 232105

\bibitem{anp04} Pasupathy A N, Bialczak R C, Martinek J, Grose J E, Donev L A K, McEuen P L and Ralph D C 2004 {\it Science} {\bf 306} 86

\bibitem{key-115} Madhavan V, Chen W, Jamneala T, Crommie M F and Wingreen N S 1998
{\it Science} {\bf 280} 567

\bibitem{hcm00} Manoharan H C, Lutz C P and Eigler D M 2000 {\it Nature} {\bf 403} 512

\bibitem{key-116} Madhavan V, Chen W, Jamneala T and Crommie F 2001 {\it Phys. Rev.} B {\bf 64}
165412

\bibitem{key-117} Knorr N, Alexander Schneider M, Lars Diekh\"{o}ner, Peter
Wahl and Klaus Kern 2002 {\it Phys. Rev. Lett.} {\bf 88} 096804

\bibitem{key-118} Wahl P, Diekh\"{o}ner L, Wittich G, Vitali L, Scheneider M A
and K. Kern 2005 {\it Phys. Rev. Lett.} {\bf 95} 166601

\bibitem{key-119} Ying-Shuang Fu, Shuai-Hua Ji, Xi Chen, Xu-Cun Ma, Rui Wu, Chen-Chen
Wang, Wen-Hui Duan, Xia-Hui Qiu, Bon Sun, Ping Zhang, Jing-Feng Jia and Qin-Kue Xue 2007 {\it Phys. Rev. Lett.} {\bf 99} 256601

\bibitem{new1} Aguiar-Hualde J M, Chiappe G, Louis E and Anda E V 2007 {\it Phys. Rev.} B {\bf 76}
155427

\bibitem{new2} Chiappe G and Louis E 2006 {\it Phys. Rev. Lett.} {\bf 97} 076806

\bibitem{ou00} \'Ujs\'aghy O, Kroha J, Szunyogh L and Zawadowski A 2000 {\it Phys. Rev. Lett.} {\bf 85} 2557

\bibitem{key-134} Patton K R, Kettemann S, Zhuravlev A and Lichtenstein A 2007 {\it Phys. Rev.} B {\bf 76} 100408(R)

\bibitem{qdefinition} The Fano parameter $q$ gives the strength of the quantum interference between different tunneling paths. It is proportional to the ratio between the tip-adatom and tip-host hopping coefficients

\bibitem{key-129} Fano U 1961 {\it Phys. Rev.} {\bf 124} 1866.

\bibitem{key-129b}Shelykh I A, Galkin N G 2005 {\it Phys. Rev.} B {\ bf 70} 05328

\bibitem{key-129c} Shelykh I A, Galkin N G, Bagraev N T 2006 {\it Phys. Rev.} B {\bf 74} 165331

\bibitem{key-129d} Moldoveanu V, Tolea M, Gudmundsson V and Manolescu A 2005 {\it Phys. Rev.} B {\bf 72} 085338

\bibitem{key-130} Frota H O and Oliveira L N 1986 {\it Phys. Rev.} B {\bf 33} 7871

\bibitem{key-132} Frota H O 1992 {\it Phys. Rev.} B {\bf 45} 1096

\bibitem{rns90} Silver R N, Gubernatis J E, Sivia D S and Jarrell M 1990 {\it Phys. Rev. Lett.} {\bf 65} 496

\bibitem{key-120} Anderson P W 1961 {\it Phys. Rev.} {\bf 124} 41

\bibitem{TKdefinition} The Kondo temperature is calculated according to equation (24) of Costi T A, Hewson A C and Zlati\'c V 1994 {\it J. Phys.: Condens. Matter} {\bf 6} 2519, $k_B T_K = \sqrt{\Gamma U/2} \textrm{exp} [\pi \varepsilon_0 (\varepsilon_0+U)/2 \Gamma U]$

\bibitem{key-41} Schiller A and Hershfield S 2000, {\it Phys. Rev.} B {\bf 61} 9036

\bibitem{key-42} Plihal M and Gadzuk J W 2001 {\it Phys. Rev.} B {\bf 63} 085404

\bibitem{Deltatio} We note that in reference \cite{key-134} the splitting $\tilde{\Delta}$ depends on $\Gamma^\uparrow_{tip}$
and not on the sum of $\Gamma^\uparrow_{tip}+\Gamma^\downarrow_{tip}$ as it was proposed in reference \cite{jm03}. In our analysis we use the proposal of reference \cite{jm03}.

\bibitem{nonequilibriumKondo} In the nonequilibrium regime, a system of a quantum dot coupled to
a left and to a right normal electrodes exhibites two Kondo resonances pinned at the Fermi levels of the leads.
See Krawiec M and Wysoki\'nski K I 2002 {\it Phys. Rev.} B {\bf 66} 165408. The influence of ferromagnetic leads
on this nonequilibrium Kondo effect was studied in reference \cite{jm03}.

\bibitem{key-46} Seridonio A C, Yoshida M and Oliveira L N {\it Preprint} cond-mat/0701529

\bibitem{key-46b} Lee W R, Kim J U and Sim H S 2008 {\it Phys. Rev.} B {\bf 77} 033305

\end{thebibliography}
\end{document}